\documentclass[useAMS,usenatbib]{mn2e}
\usepackage{graphics,epsfig}
\usepackage[dvipsnames]{color}

\def \be{\begin{equation}}
\def \ee{\end{equation}}
\def \bea{\begin{eqnarray}}
\def \eea{\end{eqnarray}}
\def\etal{{et al.\ }}

\title[Spectrum and ionization rate of low energy Galactic cosmic rays]{
Spectrum and ionization rate of low energy Galactic cosmic rays
}

\author[Biman B. Nath, Nayantara Gupta, Peter L. Biermann]
{Biman B. Nath$^1$\thanks{biman@rri.res.in}, Nayantara Gupta$^1$,
Peter L. Biermann$^{2,3,4,5}$\\
1. Raman Research Institute, Sadashiva Nagar, Bangalore 560080, India\\
2. Max Planck Institute for Radioastronomy, Auf dem H\"ugel 69, 531121,
Bonn, Germany\\
3. KIT-Karlsruhe Institute for Technology, Institute for Experimental Nuclear
Physics, 76021 Karlsruhe, Germany\\
4. Department of Physics \& Astronomy, University of Alabama, 
Tuscalosa, AL 35487,
USA\\
5. Department of Physics, University of Alabama at Huntsville, Huntsville,
AL 35899, USA\\
6. Department of Physics \& Astronomy, University of Bonn, 53115 Bonn, Germany.\\
}

\begin{document}


\maketitle

\label{firstpage}

\begin{abstract}
We consider the rate of ionization of diffuse and molecular clouds in the 
interstellar medium by Galactic cosmic rays (GCR) in order to constrain its low
energy spectrum. We extrapolate the GCR spectrum obtained from PAMELA at 
high energies 
($\ge 200$ GeV/ nucleon) and a recently derived GCR proton flux at 
$1\hbox{--}200$
GeV from observations of gamma rays from molecular clouds, and find that the 
observed average Galactic ionization rate can be reconciled with
this GCR spectrum if there is a low energy cutoff for protons 
at $10\hbox{--}100$ MeV. We also identify the flattening below a few
GeV as being due to (a) decrease of the diffusion coefficient and dominance
of convective loss at low energy and (b) the expected break in energy spectrum
for a constant spectral index in momentum. We show that the inferred CR proton
spectrum of $\Phi \propto E_{kin}^{-1.7\pm0.2}$
for $E_{kin} \le$ few GeV, is consistent with
a power-law spectrum in momentum $p^{-2.45\pm0.4}$, which we identify as the spectrum
at source. Diffusion loss at higher energies then introduces a steepening by
$E^{-\alpha}$ with $\alpha \sim 1/3$, making it consistent with high energy measurements.
\end{abstract}


\section{Introduction}
The interaction of cosmic rays (CR) with interstellar material is a powerful
probe of the spectrum, and consequently of the origin, of Galactic CR. CR
particles ionize the interstellar atoms and molecules, and the ionization
rates can be inferred from the abundance studies of different species, such
as $H_3 ^+$ (McCall et al 2003; Indriolo et al. 2007,  Goto et al. 2008). 
They also spallate
interstellar nuclei such as C, N, and O, to create light-element isotopes of
Li, Be and B (Reeves et al. 1970; Meneguzzi et al. 1971). CRs
also excite nuclear states of certain nuclei and produce gamma-ray lines,
as well as produce gamma-rays from pion decay after interacting with 
interstellar protons (Meneguzzi \& Reeves 1975; {\bf Issa \& Wolfendale 1981}). 

These interactions provide
a method of probing the spectrum of Galactic CRs at low energy where 
direct observations suffer from the effect of solar modulation.
Since these interactions depend on known 
cross-sections and an assumed extrapolation of the CR spectrum to
low energies, such studies have yielded valuable constraints on the low
energy CRs. The observed light element abundances have been
used in this regard (e.g., Meneguzzi et al. 1971; Vangioni-Flam et al. 1996;
Kneller et al. 2003), as well as gamma-ray lines (Meneguzzi \& Reeves 1975;
Ramaty et al. 1979; Cass\'e et al. 1995; Fields et al. 1996;
Tatischeff \& Kiener 2004), in addition to ionization rates (Hayakawa
et al 1961; Spitzer \& Tomasko 1968; Nath \& Biermann 1994; Webber 1998).

The observed determination of hydrogen ionization rate in diffuse interstellar
medium (ISM) $\zeta ^H$ ranges
between a few $10^{-17}$ s$^{-1}$ to a few $10^{-16}$ s$^{-1}$, based
on the abundance measurements of various species such as $OH$ and $HD$
(Black \& Dalgarno 1977; van Dishoeck \& Black 1986; Federman et al. 1996).
These inferred rates however depend strongly on various assumptions such as 
that of
the background UV radiation field. It has been shown that a cosmic ray
spectrum with a low energy cutoff $\sim 50$ MeV can explain the observed
rates (Spitzer \& Tomasko 1968; Nath \& Biermann 1994). 
For ionization
of molecular hydrogen, the standard conversion rate of $1.5 \zeta ^{H_2}
\approx 2.3 \zeta ^H$ (Glassgold \& Langer 1974) then predicted 
$\zeta ^{H_2} \le 10^{-17}$ s$^{-1}$. The observed value of $\zeta ^{H_2}$
in molecular cloud cores
however ranged between $10^{-17}\hbox{--}10^{-15}$ s$^{-1}$ (Caselli et al.
1998), with the values in high density cloud cores being $\zeta ^{H_2}$
a few times $10^{-17}$ s$^{-1}$ (van der Tak \& van Dishoeck 2000).

Recently, a high ionization rate ($\sim 4\hbox{--}6
\times 10^{-16}$ s$^{-1}$) has been determined
 in diffuse clouds through the abundance
measurement of $H_3^+$ (McCall et al 1998; McCall et al 2003; Indriolo 
et al 2007,  Goto et al. 2008). 
The simplicity of reactions involving $H_3^+$ makes 
this a robust determination.
There appears to be a range of observed molecular ionization rates from
diffuse clouds with low molecular content to that in dense clouds. Gerin
et~al. (2010) and Neufeld et~al. (2010) found an ionization rate
of $0.6\hbox{--}2.4 \times 10^{-16}$ s$^{-1}$ in diffuse clouds with
low molecular content ($H_2/H \le 10\%$),  whereas Indriolo et~al. (2010)
have found an ionization rate as large as $\sim 2 \times 10^{-15}$ s$^{-1}$
near the supernovae remnant IC 443. 
The contrast between the low ionization
rate in molecular cloud cores and the high rate in diffuse clouds can
be reconciled by either postulating that CRs are
inhibited in dense molecular clouds by some plasma processes, or that
there is an additional low energy component in the diffuse component of CRs
which cannot penetrate the dense cloud cores (Padovani et al 2009; Indriolo
et al. 2009).

Recently Neronov et al. (2012; hereafter referred to as NST12) 
have determined a spectrum of Galactic CR 
protons down to a kinetic energy of $\sim 1$ GeV, using the observed
gamma ray emission from nearby molecular clouds at the Gould's belt, and
normalizing the flux at high energy ($\ge 200$ GeV) to the flux determined
by PAMELA. They found that in order to explain the observed gamma ray
spectrum, the CR proton spectrum requires a spectral break around $9$
GeV, below which it becomes shallow. 
The low energy shape of the spectrum derived by NST12 is
interesting with regard to the above discussion of high ionization rate
$\zeta ^{H_2}$ in that the low energy flux is substantially enhanced
above the flux extrapolated from high energies. 
NST12 interpreted
the shape of the low energy spectrum as being due to the interaction
of CRs with interstellar material in the last $\sim 3 \times 10^7$ yr
while traversing a distance of $\sim 1$ kpc from Gould's belt. 
 In this paper, we suggest another possibility, that the break arises
from a combination of (a) the source spectrum being a power-law in momentum,
and (b) the dominance of convective loss over diffusion at low energies.
We then
extrapolate this spectrum down to a low energy cutoff below which
we assume the CR flux to be negligible, and compare the predicted ionization
rate with the average observed Galactic ionization rate.

Becker et al. (2011) recently considered the ionization rate by CRs in
molecular clouds (MCs) close to supernova remnants (SNRs) which are believed to
accelerate CRs. They extrapolated the theoretically predicted CR spectrum
from shock acceleration mechanism, down to MeV energies, and calculated
the predicted ionization rate as a function of a low energy cutoff, which
they used to predict abundance of (and emission line strengths from) species
such H$_2^+$. In a related paper, Schuppan  et al. (2012) considered a few
MCs associated with SNRs, and determined the CR spectrum from the observed
gamma-ray spectrum, assuming a break
at $1$ GeV and then going down to a low energy cutoff of order $30\hbox{--}100$
MeV. Then they calculated the corresponding ionization rate in the MCs, which
they found to be larger than the Galactic average in a few cases.

Our line of approach is different from these. We adopt the
spectrum of CRs determined by NST12 as the average Galactic
CR spectrum, and calculate the average ionization rate, which we compare
with the observed average ionization rate and discuss the possible low
energy cutoff for the average Galactic CR flux. Since this CR spectrum
is different from the ones adopted by recent authors, with a spectral
break at $E_b \sim 9$ GeV and with a steeper spectral index between
$E_b$ and $\sim 200$ GeV, the corresponding ionization rate cannot be
easily translated from other works. 
Our goal here is therefore to consider the average Galactic CR flux and
compute its ionization rate and compare with the ionization rate in molecular
clouds {\it not} associated with SNRs.

\section{Spectrum of low energy GCR}
The CR spectrum derived by NST12 has a spectral index $\Gamma_2
=3.03 ^{+ 0.37} _{-0.18}$ above a break at $E_b=9^{+3} _{-5}$ GeV, and a
spectral
index $\Gamma_1=1.9 ^{+0.2} _{-0.9}$. The authors state that including
a component of bremsstrahlung to the gamma-ray emission reduces the low
energy spectral index to $\Gamma_1=1.7 \pm 0.2$. The inverse Compton 
contribution is likely to be negligible if there is no star forming
region in the MC. Gabici, Aharonian \& Casanova (2009) have estimated that
inverse Compton
 loss becomes significant only if there is an OB association with a total
output of $\sim 4 \times 10^{33}$ erg/s located in the MC. In this case
of MCs considered by NST12, there are no SNRs associated with them, and
we assume that inverse Compton loss is not significant.
After inferring the spectral indices from gamma ray emission, 
NST12 normalizes the CR spectrum with the high energy measurement $\ge 200$
GeV with PAMELA (Adrinai et al. 2011). 

We note that the CR spectrum inferred at
 high energy ($E_{kin} \ge 10 $ GeV) by NST12, with
 spectral index of $3.03 ^{+ 0.37} _{-0.18}$ is 
consistent with {\bf or close to} that measured by different experiments.
For example, PAMELA
finds a spectral index of $\sim 2.85$ for protons in the range of 
$80\hbox{--}230$ GeV (Adriani et al. 2011). {\bf However} 
AMS measured an index of $2.78\pm0.009$
at $10\hbox{--}200$ GeV (Aguilar et al. 2002), BESS an index of
$2.732\pm 0.011$ at $30\hbox{--}{\rm few}$ hundred GeV (Haino et al. 2004), and 
CREAM measured an index of $\sim 2.66\pm0.02$
between $2.5\hbox{--}250$ TeV (Yoon et al. 2011). 

The expected spectrum of CRs from diffusive shock acceleration is a power-law
in momentum ($p$) (e.g, Bell 1978; Drury 83; 
Blandford \& Eichler 1987 and references
therein).
Although at high energies the corresponding spectral index in energy is 
the same as that for momentum, since $E \propto pc$, there is expected to
be a break in the energy spectrum at a few GeV, with a different spectral
index for the low energy CRs, for which one should use 
$E_{kin}=\sqrt{p^2 c^2 + m^2 c^4} -mc^2$ instead of the approximation of $E\sim pc$. 
We show in Figure 1 that a spectrum of $\Phi (p) \propto p^{-2.45}$
shows a break in the spectrum in kinetic energy at a few GeV, with the
low energy spectral index being $\sim 1.7$, and recovering the index of 
$2.45$ at higher energies. {\bf The figure shows that the energy spectrum
changes at $\sim \log _{10} (E_{kin}/m_pc^2) \le 0.6$, or $E_{kin} \le
3.7$ GeV, that lies in the range of measurement of NST12 ($E_{kin} \ge 2$ GeV),
and is comparable to the lower (1$-\sigma$) error margin of the break energy 
$E_b$ ($9_{-5}$ GeV).
}
Our choice of the spectral index is aimed
at recovering the inferred index of $1.7$ at low energies by NST12.
The uncertainty in the inferred spectral index ($-1.7\pm0.2$) translates
to an uncertainty in the momentum spectral index of $-2.45\pm0.4$.

We also note that the diffusion coefficient of CRs in the ISM is thought
to be energy dependent, and this energy dependence steepens the observed
CR spectrum compared to the source spectrum. 
 For protons ($Z=1$), the diffusion coefficient at high energies scales
as
$D \propto \upsilon \; p^{\alpha}
$, where $\alpha$ depends on the nature of turbulences which scatter
the CRs, $\alpha \sim 1/3$ for Kolmogorov-type spectrum of turbulence
(e.g., see \S 2 in Ptuskin \etal 2006).
At low energies, the diffusion coefficient becomes independent
of energy below an energy of $\sim 3$ GeV (for protons, with $Z=1$), as 
suggested by the observed ratios secondary to primary nuclei
(Ptuskin \etal 2006).
However, as Biermann et al. (2001) have suggested, convective transport is
likely to become more dominant than diffusive ones since it is faster,
 at energies $\le $
few GeV/n. 
{\bf In other words,
the mode of cosmic ray transport likely changes at $E_{kin} \sim $ few GeV/n,
with convection dominating at low energies and diffusion at high energies.
Therefore, 
if we identify $p^{-2.45\pm0.4}$ as the source spectrum, then
(a) at low energies (a few GeV/n), diffusion loss does not steepen the spectrum,
and we get a spectral index of $-1.7\pm0.2$ in kinetic energy, and
(b) the high energy spectrum will steepen by $d \gamma\sim
1/3$, rendering an index 
$\sim -2.8\pm0.4$,  
consistent with the high energy measurements
as mentioned above, especially the AMS data.}
Also, the inferred spectral index at source is consistent with 
theoretical predictions (see, e.g., Biermann 1993).

NST12 proposed that the break below $\sim 10$ GeV arises either because of
p-p interactions with ISM or scattering from magnetic inhomogeneities which
may be dominated by clouds of a certain size ($\sim 1$ AU). However, from
the p-p interaction argument, the time scale is $\sim 3 \times
10^7$ yr, much larger than the typical residence time scales of GCR inferred
from other considerations.

In view of the {\bf arguments} presented above, which makes two
reasonable assumptions: (a) that the source spectrum of CRs is a power-law
in momentum and that (b) the diffusion coefficient becomes energy independent
and/or becomes sub-dominant compared to convective transport, below $\sim 5$
GeV, we find that
the NST12 spectrum can be explained without recourse to any 
additional 
processes, {\it although one cannot rule out other processes contributing
towards the spectral shape}.

\begin{figure}
\centerline{
\epsfxsize=0.4\textwidth
\epsfbox{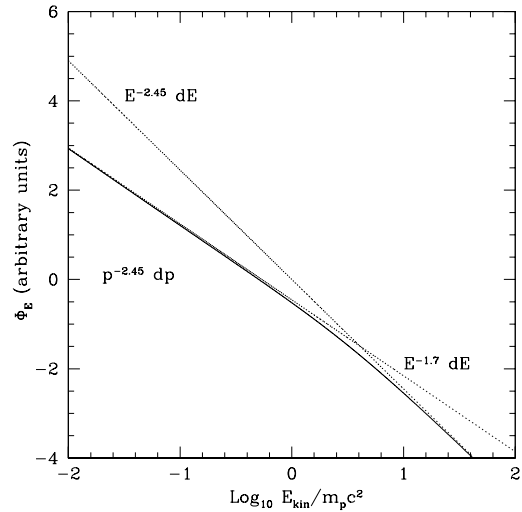}
}
{\vskip-3mm}
\caption{
CR proton 
energy spectra are shown with fluxes in arbitrary units, against kinetic
energy in the units of rest mass energy. The solid lines 
shows a spectrum with constant power-law in momentum (with an index $-2.45$),
and the dotted lines show the asymptotic behaviour at low and high energy
ends.
}
\end{figure}

\section{Ionization rate}
Next, we wish to constrain the NST12 spectrum by determining a low energy
cutoff below $\sim 1$ GeV after comparing with the observed molecular
ionization rate.

Normalizing the spectrum derived by NST12 to the PAMELA
flux above $200$ GeV, and given the observed PAMELA flux at the energy
bin centering $213.23$ GeV to be $(4.17 \pm 0.08 \pm 0.2) \times
10^{-7}$ cm$^{-2}$ sr$^{-1}$ s$^{-1}$ GeV$^{-1}$ (Adriani et al. 2011),
 we adopt the following spectrum for the diffuse
Galactic CR protons:
\bea
&& {dN_p \over dE_p} \nonumber\\
&=& 4.17 \times 10^{-7} {\rm cm}^{-2} \, {\rm s}^{-1} \,
{\rm sr}^{-1} \, {\rm GeV}^{-1} \Bigl ( {E_p \over 213.23 \, {\rm GeV}}
\Bigr )^{\Gamma_2} \,, \nonumber\\
&& \qquad\qquad\qquad E_p \ge E_b \nonumber\\
&=& 4.17 \times 10^{-7} {\rm cm}^{-2} \, {\rm s}^{-1} \,
{\rm sr}^{-1} \, {\rm GeV}^{-1} \nonumber\\
&& \times \Bigl ( {E_b \over 213.23 \, {\rm GeV}}
\Bigr )^{\Gamma_2} \, \Bigl ( {E_p \over E_b} \Bigr )^{-\Gamma_1} \,,
\qquad E_p  < E_b \nonumber\\
&& 0 \,, E_p \le E_{low}
\eea

We follow  Padovani et al. (2009) in our calculation of the ionization
rate of molecular and atomic hydrogen. Only the direct ionization by primary
protons is significant and dominates over other processes such as ionization
by CR electrons and electron capture by CR protons by more than an order of
magnitude. The ionization rate of $H_2$ by primary protons is,
\be
\zeta ^{H_2}= \int _{E_{low}} ^\infty {dN_p \over dE_p} \, \sigma_p ^{ion}
(E_p) \, dE_p \,,
\ee
where ${dN_p \over dE_p}$ is the CR proton spectrum at a given kinetic
energy $E_p$ and $E_{low}$ is the low energy limit of $E_p$. The
cross-section $\sigma_p ^{ion}$ is given in Padovani et al. (2009).
We also multiply this ionization rate by a factor $5/3$ in order to take
into account the ionization by secondary electrons (Spitzer \& Tomasko 1968,
Dalgarno \& Griffing 1958).

The helium nuclei associated with the CR protons will enhance the ionization
of the ISM molecules. In order to assess the enhancement factor introduced
by the addition of helium nuclei, we use the ratio of protons to helium
nuclei as a function of kinetic energy per nucleon as determined by PAMELA
down to a $0.4$ GeV/n, and use the proton spectrum of NST12
down to the same value of kinetic energy. We found that the ratio
of proton
to helium flux in PAMELA data (Adriani et al. (2011) can be fit to an accuracy
of $\sim 10 \%$ to the lowest kinetic energy (per nucleon) bin in their tables
S3 and S4, by the following expression:
\begin{equation}
{\Phi (p) \over \Phi (He)} \approx 4.47+12.68 \, \tanh (\epsilon /2.247 \, 
{\rm GeV}) \,,
\end{equation}
where $\epsilon=E_k/A$ is the kinetic energy per nucleon. This ratio 
becomes constant above a value of $\epsilon \ge 20$ GeV/n, and decreases
at lower energies. 

We find that the ionization rate increases by a factor $1.48$ if helium
nuclei are included, using the ratio of $p/He$ along with the NST12
spectrum, for a low energy cutoff of $0.4$ GeV/n.
We note that the $p/He$ ratio as determined by PAMELA may however
 be affected by solar
modulation, and therefore the enhancement of ionization may also be affected.
If the low energy is shifted to $1$ GeV/n, then the ionization rate increases
by a factor $1.33$, and for a cutoff at $10$ GeV/n, the corresponding factor
is $1.23$. We note that these factors are somewhat lower than the standard
factor of $1.8$ used in literature (e.g., Rimmer et~al. 2011).
 Keeping this uncertainty in mind, we multiply the ionization
rate of protons by a factor $1.5$ below in order to take helium nuclei
into account.

\begin{figure}
\centerline{
\epsfxsize=0.4\textwidth
\epsfbox{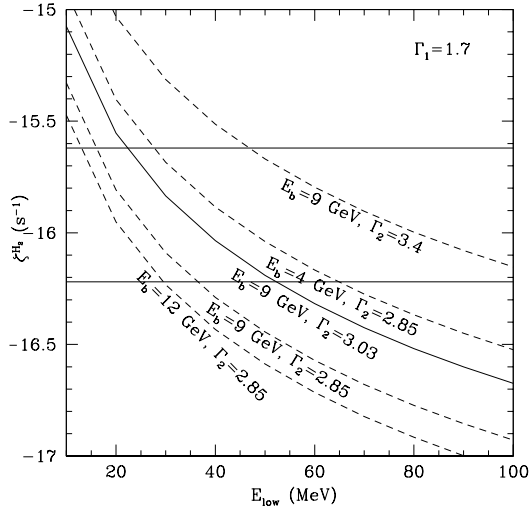}
}
{\vskip-3mm}
\caption{
The molecular ionization rate $\zeta ^{H_2}$ is calculated for the
spectrum mentioned above, as a function of the low energy cutoff $E_{low}$,
for a few combinations of the parameters $E_b$ and $\Gamma_2$.  The low
energy spectral
index $\Gamma_1$ has been fixed at $1.7$.
}
\end{figure}

We show in {\bf Figures 2 and 3} 
the corresponding ionization rate $\zeta ^{H_2}$
as a function of the low energy cutoff, $E_{low}$, and also show
the observed {\bf range} of $\sim 0.6\hbox{--}2.4 \times 10^{-16}$ s$^{-1}$,
{\bf for the rates in diffuse clouds with low molecular content
(Gerin \etal (2010); Neufeld \etal (2010)). We choose this range
since the propagation effects are least likely to affect in the
case of low density diffuse clouds.} The curves show
the ionization rates
for different combinations of $E_b$ and $\Gamma_2$, {\bf but with 
$\Gamma_1=1.7$ in Figure 2 and $\Gamma_1=1.9$ in Figure 3.
We note that NST12 derived a value of $\Gamma_1=1.7$ in the case
of maximal bremsstrahlung component, and provides
an extreme limit.}

\begin{figure}
\centerline{
\epsfxsize=0.4\textwidth
\epsfbox{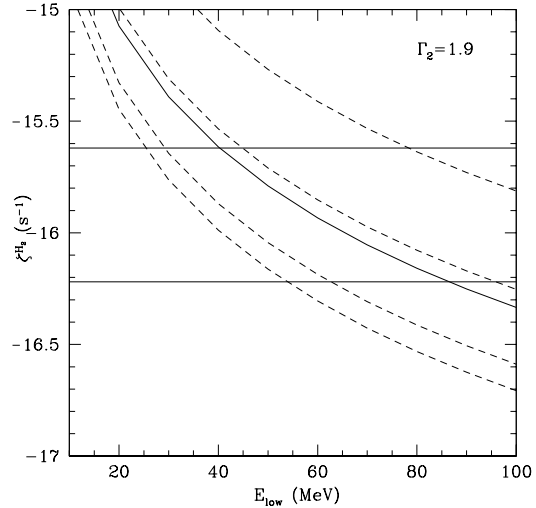}
}
{\vskip-3mm}
\caption{
Same as Figure 2 except for $\Gamma_2$ being fixed at $1.9$.
}
\end{figure}

{\bf The curves in Figure 2 and 3 show that the 
ionization rates are somewhat lower in the case of $\Gamma_1=1.7$ than
for $\Gamma_1=1.9$. Also, we find that the
corresponding values of $E_{low}$ ranges between $\sim 10\hbox{--}100$ MeV,
for the average $\xi^{H_2} \sim 0.6\hbox{--}2.4 \times 10^{-16}$ s$^{-1}$
for diffuse clouds with low molecular content. 
}
We note that
Schuppan et al. (2012) in their determination of the CR spectrum for MCs
associated with SNRs, did not add the bremsstrahlung component, and therefore
their predicted high ionization rate is likely to be an overestimate.

\section{Discussion}
The  low energy cutoff $E_{low}$ can arise from a number of effects, and we
would like to discuss these possibilities here.

\subsection{Propagation effects}
The low energy cutoff $E_{low}$ is related to
the quantity of matter (grammage) the CRs traverse on average in the Galaxy,
and the ionization losses suffered by the protons below $E_p \le E_{low}$.
The ionization loss for protons traversing the ISM is given by eqn 4.22 of
Mannheim \& Schlickeiser 1994,
\be
{dE_p\over dt}(\beta)=4.96 \times 10^{-19}\,{\rm erg~s^{-1}}~(n_e/ {\rm cm^{-3}})
\,{ \beta^2 \over \beta^3 +x_m^3},
\label{eq:loss}
\ee
where $x_m=0.0286 (T/2 \times 10^6 \, {\rm K})^{1/2}$, $T$ and $n_e$ are the 
ambient
(electron) temperature and density, and $\beta=v/c$. The ionization loss
increases rapidly with decreasing proton energy.

For low temperature ISM gas,
this ionization loss of protons
can also be expressed in terms of the grammage ($\sim n m_p\beta ct$)
as ${\Delta E_p \over E_p} \sim {6.7 \times 10^{-2}  \over \beta ^2 (\gamma
-1) } \Bigl ( {{\rm grammage} \over 10 \, {\rm g} \, {\rm cm}^{-2}}
\Bigr )$, where $\gamma$ is the Lorentz factor. 
If we identify the cutoff energy $E_{low}$ as due to ionization loss,
then the inferred grammage for $E_{low}=10 (50)$ MeV
is $\sim 0.03 (0.80)$ g cm$^{-2}$. The grammage for a $100$ MeV proton
is $\sim 3$ gm cm$^{-2}$. 

The average grammage inferred for GeV cosmic rays in our Galaxy
is $\sim 10$ g cm$^{-2}$ (e.g., Brunetti \&
Codino 2000). This grammage is believed to be smaller for lower energy
protons, and go down to $\sim 3\pm 1$ gm cm$^{-2}$, for $E_{kin} \le 100$
MeV, and also decrease with higher energies (e.g., see Fig 7 in
Garcia-Munoz et~al (1987)). Interestingly, this estimate of grammage
inferred from the secondary/primary ratio is  close to the
grammage inferred from the interpretation of low energy cutoff of 
$10\hbox{--}100$ MeV from ionization rate, as 
discussed above. This is our primary result, that using the NST12 spectrum
allows us to identify the low energy cutoff with ionization rate, 
in the sense that the grammage inferred from B/C ratio at this energy
coincides with the grammage expected from ionization loss. That this near
equality is 
difficult to achieve with other CR spectrum is evident. If we were to 
assume a CR spectrum without a flattening
down to a low energy cutoff, with a spectral index $\sim 2.85\hbox{--}3.$, 
then the Galactic average ionization rate would
constrain the low energy cutoff to $\sim 250\hbox{--}350$  MeV. However,
if this were to be identified as arising from ionization loss, then the
corresponding grammage would be $15\hbox{--}30$ gm cm$^{-2}$, much
larger than even the peak grammage at GeV scale. Therefore our result
shows that NST12 spectrum is consistent with the average
Galactic ionization rate with a corresponding grammage of CR protons at
low energy arising from ionization loss.

Previous workers (Indriolo et al. (2009); Becker et al.
(2011)) have used a low energy cutoff of $\sim 2$ MeV for diffuse clouds,
and $10$ MeV for dense clouds. It has been argued that the range of a 1 MeV
proton is $8.5 \times 10^{-4}$ g cm$^{-2}$, corresponding to a column
density of $\sim 5 \times 10^{20}$ H cm$^{-2}$. Indriolo et al. (2009)
considered a low energy cutoff of $2$ MeV for diffuse clouds with
density $\le 10^3$ cm$^{-3}$, and Becker et al. (2011) assumed
a low energy cutoff of $10$ MeV for molecular clouds near SNRs, with density
$\ge 10^3$ cm$^{-3}$.
Here also, the idea has been to find a consistency between the CR spectrum,
the low energy cutoff and the ionization rate. However, 
the calculations mentioned
here have either assumed a CR spectrum (and found the ionization rate) or
found a CR spectrum for a given ionization rate. At times finding a convergence
has been difficult. For example,
Indriolo et~al (2009) found
that their model CR spectrum that took propagation of CRs in the Milky Way
produced a lower ionization rate than observed, and they suggested the
presence of an additional component of low energy CRs. 
Our result here shows
that with the NST12
spectrum, a convergence of low energy cutoff, grammage and ionization rate
is possible.

Also, our results show that even for diffuse clouds a low energy cutoff below 10
MeV is inconsistent with the Galactic average ionization rate, and that
the low energy cutoff lies in the range of $10\hbox{--}100$ MeV. Therefore,
the cutoff in dense molecular clouds is likely to be larger than this value,
and our results suggest that the ionization rate in molecular clouds with SNRs,
as in Becker et al. (2011), is likely to be an overestimate because of the
assumption of a low value of cutoff.

Interestingly, in the model of Biermann et~al. (2001), the CR protons from
SNR shocks in ISM are expected to have a grammage of $\sim 1$ gm cm$^{-2}$,
similar to the convergence value found in our analysis. CRs from massive
star supernovae are believed to consist of mainly heavy nuclei and their 
spallation products, while supernovae of moderate mass stars produce
predominantly CR protons. The secondary
 CRs produced in the spallation 
interactions of massive star supernovae with stellar wind material, are 
expected to
have a much larger grammage. Therefore our results support the acceleration
of CR protons
in SNR shocks in ISM.

\subsection{Magnetic field effects}
Cesarsky \& V\"olk (1978) proposed that CR streaming instability in the
presence of magnetic field in MCs would excite Alfv\'en waves which would
scatter off low energy CRs (Skilling \& Strong 1976). 
They estimated that this would screen the MCs
from Galactic low energy CRs, and that the low energy cutoff would be
related to the cloud size ($R$), particle density $n$ and magnetic field
$B$, as
\begin{equation}
E \sim 50 \, {\rm MeV} \, (R/8.2 \, {\rm pc})^{0.5} \, (n/2 \times 10^3 \, {\rm cm}^{-3})^{0.5}
\, (B / 50 \mu {\rm G})^{-0.5} \,.
\end{equation}
This shows that for a typical MC magnetic field of a few $\mu$G (Curran
\& Chrysostomou 2007), the
low energy cutoff can be as large as $\ge 100$ MeV. However, if we turn the
argument around, then given the typical values of $E_{low}$ derived above
for CRs inside MCs imply that the ability of magnetic fields to screen
low energy CRs in MCs is smaller than previously thought, since it is clear
that in order to sustain the typical average ionization rate in MCs,
$E_{low} \le 100$ MeV.

\section{Summary}
 In summary, 
we have studied the Galactic CR proton spectrum recently inferred
from gamma ray emission from molecular clouds near Gould's belt. 
We have identified the 
inferred break of the energy spectrum
at a few GeV as a consequence of (a)  the source CR spectrum
being a power-law in momentum, of $p ^{-2.45}$, and (b) the diffusion
coefficient for protons being independent of energy below a few GeV. 
We have shown that this renders the CR spectrum consistent with that
observed at high energies.
We have also derived a low energy cut-off
for the CR spectrum, after comparing with the observed Galactic ionization
rate of molecular hydrogen (since the ionization is dominated by
protons), and found its values to lie in the range of 
$10\hbox{--}100$ 
MeV.
We have then shown that the NST12 spectrum is consistent with three independent phenomena:
(a) ionization loss that corresponds to a particular value of grammage,
(b) grammage as inferred from other means, and (c) the average Galactic
 ionization  rate. 


\medskip
We thank the anonymous referees for their comments that helped to improve
the paper.

\end{document}